# Application of an Equilibrium Phase (EP) Spray Model to Multi-component Gasoline Direct Injection

Zongyu Yue[1*], Rolf D. Reitz[2]

[1]Argonne National Laboratory, USA   [2]University of Wisconsin-Madison, USA

**Abstract**

An Equilibrium Phase (EP) spray model has been recently proposed for modeling high-pressure diesel fuel injection, which is based on jet theory and a phase equilibrium assumption. In this approach, the non-equilibrium processes of drop breakup, collision and surface vaporization are neglected, assuming that spray vaporization is a mixing-controlled equilibrium process. A liquid-jet model and a gas-jet model are also introduced to improve grid-independency. In the current study, the EP model is applied in simulations of multi-hole gasoline direct injection (GDI). The model is validated at ambient densities from 3 to 9 kg/m$^3$ and ambient temperatures from 400 K to 900 K, for two different GDI injectors, i.e. the Engine Combustion Network (ECN) Spray G injector and a GM injector. Iso-octane is used as the surrogate in consistency with available experiments. The results show good agreement in terms of liquid/vapor penetration, shape of the vapor envelope, and the velocity evolution along the injector centerline. Then, a 10-component gasoline surrogate fuel is used to demonstrate the capability of this model for multi-component spray simulations, which is essential for engine combustion and emission predictions.

Keywords: GDI, ECN, Spray G, CFD, phase equilibrium

## Introduction

Gasoline direct injection (GDI) in spark-ignition engines has been a trend in the light-duty market for the past decade in order to further improve fuel economy and reduce $CO_2$ emission. Compared to the port fuel injection system, the GDI system offers precise fuel delivery, less cycle-to-cycle variation (CCV), improved fuel economy by engine knock mitigation and allowing higher compression ratios, and the potential for unthrottled, stratified lean combustion [1]. Fuel injection is one of the critical processes in a GDI engine, especially for spray-guided combustion systems, which make use of the injection process to form a stably ignitable fuel/air mixture around the spark plug. The multi-hole GDI nozzle has the advantage of providing stable spray structure and it is less sensitive to operating conditions compared to pressure-swirl atomizers [2]. Despite the similarity to multi-hole diesel injectors, multi-hole gasoline injectors usually feature a narrower drill angle, smaller length/diameter (L/D) ratios, a step-hole design, and a relatively lower fuel pressure (10-40 MPa), with a more volatile, less viscous fuel being injected into a cooler and lower density chamber.

CFD modeling of the fuel injection process has long been a focus for diesel applications [3]-[5], and has been adapted for GDI spray simulation as well [6]. Briefly, the liquid is injected as discrete 'blobs' with an initial size of the effective nozzle radius to represent the intact liquid core, following which the processes of primary/secondary breakup, collisions and vaporization are also modeled. Since gasoline and diesel fuels are complex mixtures that consist of a number of paraffins, cycloalkanes and olefins, etc., a continuous composition model [7] or a discrete component model [8] with a suitable multi-component surrogate is used to model the liquid vaporization.

While much effort has been focused on modeling liquid breakup and droplet vaporization, experimental studies have revealed the mixing-controlled characteristics of high pressure fuel injection under both diesel and gasoline engine relevant conditions [9][10]. In this scenario, the liquid jet is analogous to a turbulent gas jet and the vaporization process is controlled by the entrainment rate of hot ambient gas, rather than by liquid breakup and mass transfer rates at droplet surfaces. An equilibrium phase (EP) spray model [11] has been recently proposed for engine CFD simulations, which is based on this mixing-controlled jet theory and the assumption of local phase equilibrium. This model has been applied to simulate diesel-type fuel injection and has shown excellent predictions in terms of transient vapor/liquid penetrations and spatial distributions of mixture fraction under a wide range of conditions without the need of model constant tuning. Moreover, the model shows better computational efficiency than the classical spray modeling approach since the non-equilibrium processes of drop breakup, collision, and vaporization are not modeled.

In this study, the EP spray model is adopted for GDI spray simulations. The Engine Combustion Network (ECN) Spray G and GM GDI iso-octane spray experiments are used for model validation. Further, a 10-component gasoline surrogate is used to demonstrate the capability of multi-component spray simulations with the EP model.

**Model description**

The EP spray model is discussed in detail in Reference [11], and is only briefly described here. The approach emphasizes the role of the mixing process by introducing an Eulerian liquid phase, solving for the local phase equilibrium, while neglecting non-equilibrium processes of drop breakup, collision and finite-rate interfacial vaporization. The fuel is injected into the computational domain as Lagrangian parcels in order to avoid the need to resolve the spray at the nozzle scale. The liquid fuel is gradually released from the injected Lagrangian parcels to the Eulerian liquid as it moving away from the nozzle exit. The governing equation for the Eulerian liquid reads,

$$\frac{\partial \rho_l}{\partial t} + \nabla \cdot (\rho_l \mathbf{u}) = \nabla \cdot \left[\rho D \nabla \left(\frac{\rho_l}{\rho}\right)\right] + \dot{S}_{EP} + \dot{S}_{release} \quad (1)$$

$\rho_l$ is the mass density of the Eulerian liquid. $\dot{S}_{EP}$ is the source term due to phase change between the Eulerian liquid and vapor based on local phase equilibrium, which is limited by entrainment and mixing with the ambient hot gas. $\dot{S}_{release}$ is a source term that describes the transition from Lagrangian drops to the Eulerian liquid, which is governed by the Liquid-Jet model,

$$\dot{S}_{release} = \left(m_p - m_{p,i} \cdot (1 - \gamma(x))\right)/dt \quad (2)$$

$$\gamma(x) = \frac{\dot{m}_a(x)}{\dot{m}_a(L)} = \frac{\sqrt{1+16\tilde{x}^2}-1}{\sqrt{1+16(L/x^+)^2}-1} \quad (3)$$

$m_p$ is the mass of the Lagrangian liquid drop parcel, and $m_{p,i}$ is the mass of the same parcel when initially injected. $\dot{m}_a(x)$ is the mass flow rate of entrained ambient gas as a function of axial distance, $x$. $\tilde{x} = x/x^+$ is the axial distance normalized by $x^+ = \sqrt{\rho_f/\rho_a} \cdot \sqrt{C_a} \cdot d/\tan(\theta/2)$. $\rho_f$ is the fuel density, $\rho_a$ is the ambient gas density, $d$ is the nozzle diameter, and $C_a$ is the area-contraction coefficient of the nozzle and is set to be 0.8 in this study. The function $\gamma(x)$ is the ratio of the mass flow rate of the entrained ambient gas at any axial location and the one at that liquid length tip required to totally vaporize the fuel. In a mixing-controlled regime, $\gamma(x)$ also represents the possible upper bound of the proportion of fuel being vaporized at any axial location. $L$ is the liquid length and estimated by a scaling law,

$$L = C_L \cdot \sqrt{\frac{\rho_f}{\rho_a}} \cdot \frac{\sqrt{C_a} \cdot d}{\tan(\theta/2)} \sqrt{\left(\frac{2}{B(T_a,P_a,T_f)}+1\right)^2 - 1} \quad (4)$$

$C_L = 0.62$ is a model constant [9]. The term $B(T_a, P_a, T_f)$ is the mass ratio of the injected fuel and entrained ambient gas in a saturated vapor condition, which is a function of ambient temperature ($T_a$), pressure ($P_a$) and fuel temperature ($T_f$). The cone angle $\theta$ is modeled as function of gas/liquid density ratio, $\frac{\rho_f}{\rho_a}$, Reynolds number, $Re_f$ and Weber number, $We_f$, according to the aerodynamic surface wave theory of jet breakup [12],

$$\tan \theta/2 = C_\theta 4\pi \sqrt{\frac{\rho_a}{\rho_f}} f(\frac{\rho_f}{\rho_a}(\frac{Re_f}{We_f})^2) \quad (5)$$

$C_\theta$ is a model constant that depends on injector configuration and nozzle flow.

The EP model was implemented in the KIVA-3vr2 code [13], with several sub-models improved for engine combustion simulations. A Gas-jet model [14][15] is used to improve grid-independency by calculating the liquid/gas momentum coupling with a sub-grid velocity from turbulent gas-jet theory. A generalized-RNG k-ε turbulence model [16][17] is used for turbulence modeling. An advanced phase equilibrium solver [18] that considers phase stability and flash calculations is employed. The Peng-Robinson equation of state [19] is applied for both the phase equilibrium solver and the gaseous phase Pressure-Volume-Temperature relationship to account for real gas effects under high pressures and high temperatures [20].

**Results and discussion**

*ECN Spray G*

The model is first validated at the standard ECN Spray G condition [21]. The Spray G injector has eight holes with a stepped-hole structure and a drill angle of 37˚. However, the measured value of 33˚ [22] is used for the plume direction in this study. The nozzle diameter is 165 μm and the length/diameter ratio is 1.03. Iso-octane was used as a surrogate for gasoline in the experiments and in the model. The injected fuel temperature was 363 K and the injection pressure was held at 20 MPa. The duration of the injection was 780 μs and the measured rate of injection is shown in Figure 1. The constant volume spray chamber is maintained at 573 K, 3.5 kg/m$^3$ with nitrogen. For the simulation setup, the computational domain is a cylinder with diameter of 10 cm and height of 10 cm. A 45˚ sector mesh is used to simulate one spray plume. In this case, plume-to-plume interaction is simplified assuming each plume is identical. The mesh resolution is 0.5 mm * 0.5 mm * 4.5˚ near the nozzle, and transitions to 2.75 mm * 2.75 mm * 4.5˚ at the far sides. Due to lack of experimental measurement of the cone angle, $C_\theta$ is determined to be 1.2 based on a best match of jet penetrations, and the cone angle prediction is also shown in Figure 1. Since the cone angle is modeled as function of Reynolds number and Weber number that depend on the instantaneous injection velocity, the transient periods are captured at the start and end of injection. A cone angle of 25 degrees is predicted for the quasi-steady period.

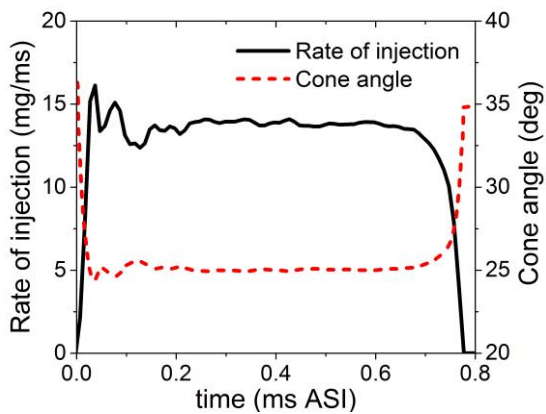

*Figure 1 Rate of injection measurement and spray cone angle prediction.*

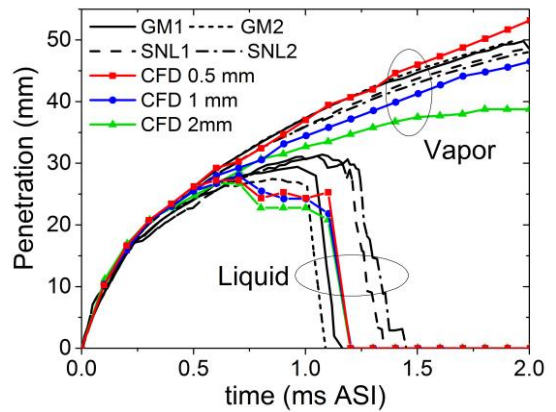

*Figure 2 Liquid and vapor penetrations for ECN Spray G injector.*

Schlieren and Mie-scatter imaging were used in the experiments to track the envelopes of the vapor jets and liquid sprays, respectively. The liquid and vapor penetrations were evaluated for the entire spray rather than for a single plume, and the distance from the nozzle exit to the jet tip was measured along the direction of the injector axis. In the CFD simulation, the liquid penetration length is defined as the maximum distance from the nozzle outlet to the farthest axial position for a 0.01% liquid volume fraction. The vapor penetration is determined as the maximum distance from the nozzle outlet to the tip region where the fuel vapor mass fraction is 0.1%. Figure 2 shows the spray penetration predictions in comparison with available experimental measurements. Several experimental measurements made by different institutes are plotted with black lines [21]. It is seen that vapor penetration measurements are relatively consistent, while the liquid penetration shows more variations, especially for the period after the end of injection. The red lines are the CFD predictions with 0.5 mm mesh resolution, and show good agreement with the measurements for both the vapor and liquid penetrations. A discrepancy in the liquid length is seen after the end of injection, which can be partly attributed to the larger uncertainty in the measurements. Coarser meshes of 1 mm and 2 mm resolution were also tested in the simulations, and are shown as blue and green lines in Figure 2. It is seen that the liquid length is less sensitive to the mesh resolution, due to the use of the Liquid-Jet and Gas-Jet models that enable accurate liquid/gas momentum exchange predictions that are independent of mesh resolution. However, considerable discrepancy is seen in the vapor penetration in that the coarser mesh has a slower penetration rate at later stages. Despite the accurate momentum exchange with the liquid, fine spatial meshing in the near-nozzle region is still required to correctly resolve the velocity profile, which can also be illustrated by examining the axial velocity.

The evolution of the axial velocity component along the injector centerline 15 mm downstream of the injector tip is shown in Figure 3. The black line and shadow areas are the ensemble average and standard deviation of PIV measurements, which present a recirculation period during the injection where the flow is moving upward to the injector with a negative axial velocity. The recirculation velocity starts to decelerate at 0.75 ms after the start of injection (ASI), slightly after the point when the rate of injection starts to drop. With the plume slowing down and widening and collapsing along the centerline after the end of injection, the flow at the measured location reverses to be directed downward with a positive axial velocity. This transient velocity profile is well predicted by the current model using the 0.5 mm mesh resolution, in terms of velocity magnitude and phasing with a slight delay in the plume arrival time, as shown by the red line in Figure 3. The blue line and green line show the velocity predictions with coarse meshes, which under-predict the recirculation velocity and significantly over-predict the downward-directed velocity with an early reversal time. Figure 4 gives the distribution of the velocity field for 0.5 mm and 2 mm mesh resolution, respectively. The red spheres represent the injected Lagrangian parcels. Black arrows indicate velocity vectors, and the size of the arrow tip is proportional to the velocity magnitude. The location of the hole exit is 0.79 mm in the radial direction from the injector axis, which means that with a mesh resolution larger than 0.79 mm, the hole exit is within the cell that is adjacent to the axis of symmetry of the sector domain. This results in excessive momentum transfer to the gas phase along the injector centerline because the radial velocity along the axis of symmetry is zero. The over-predicted axial velocity suppresses the recirculation flow and leads to an unrealistic collapse along the axis after the end of injection. The dispersed velocity profile seen in Figure 4 (b) also results in a slow penetration rate, as found in Figure 2. This grid dependency has not been seen when applying the present EP spray model to diesel spray simulations, which is likely due to the wider hole drill angle for diesel injectors.

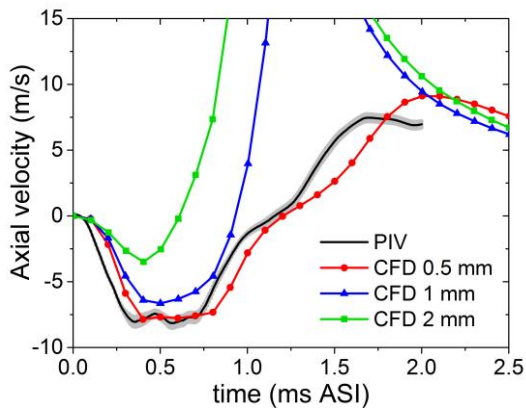
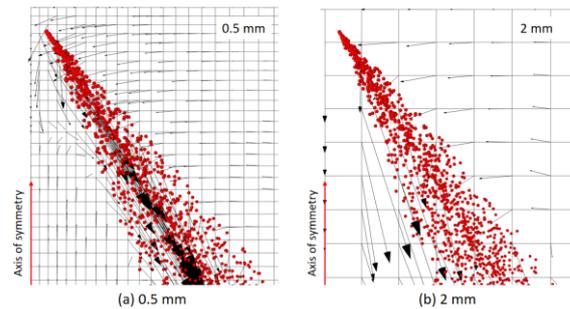

*Figure 3 Temporal axial velocity on the centerline at 15 mm downstream of the injector tip.*

*Figure 4 Velocity field for 0.5 mm and 2 mm mesh resolution*

### GM GDI

Optimization of the injection strategy in engines using CFD tools requires the spray model to be predictive across a wide range of in-cylinder conditions, covering from cold start, early injection during the intake stroke, to late injection for mixture stratification. Thus, validation at multiple conditions is more challenging and is a better demonstration that the key underlying physics are being correctly modeled, compared to validation at a single condition where a good result can usually be achieved by optimizing model constants. In this section, the EP model is validated against experiments by Parrish at GM [10], at 18 operating conditions, as shown in Figure 5. Three ambient densities were targeted, with temperatures varying from 400 K to 900 K. The GDI nozzle used here is also an eight-hole counter-bore injector, similar to the ECN Spray G injector, but has a larger L/D ratio of 1.8 and narrower plume direction of 25°. Injection pressure was maintained at 20 MPa with 766 μs duration. The fuel temperature was 363 K. Since $C_\theta$ is negatively correlated with L/D ratio [12], a smaller value of 0.6 is used for $C_\theta$ to best match the liquid length at 6 kg/m$^3$, 700 K. All the other model details were kept the same for the Spray G simulation, and also unchanged across the different conditions in this section.

The predictions of the liquid and vapor envelopes are compared with the experimental measurements for 700 K at three ambient densities, as shown in Figure 6. For each operating condition, the results are compared at 0.5 ms, 1.0 ms and 1.5 ms ASI, with the experimental results on the left and the simulation results on the right. For the experimental line-of-sight measurements, red and green lines outline the liquid and vapor envelopes, respectively, and the black dashed lines mark the measured spray included angle. For the CFD predictions white clouds represent iso-volumes where the vapor mass fraction is larger than 0.1%, and red clouds represent iso-volumes where the liquid volume fraction is larger than 0.01%. The result of a simulated single plume in a sector mesh is replicated and rotated in order to compare with the experimental results. The general trend is that both spray penetration and liquid residence time decrease with increasing ambient density, which is found in both the experimental results and the CFD predictions. The included angle and penetration rate of the vapor envelopes seen in the experiments are well captured by the simulations at each transient time for each operating condition. Similar accuracy can be found in the simulation results for 500 K and 900 K cases.

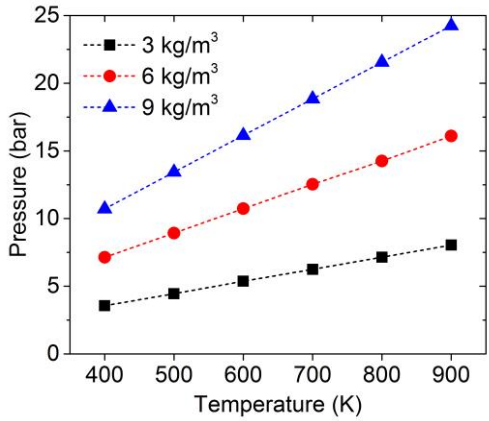

*Figure 5 Operating conditions for GM GDI sprays. Black squares: ambient density at 3 kg/m³; Red circles: ambient density at 6 kg/m³; blue triangles: 9 kg/m³.*

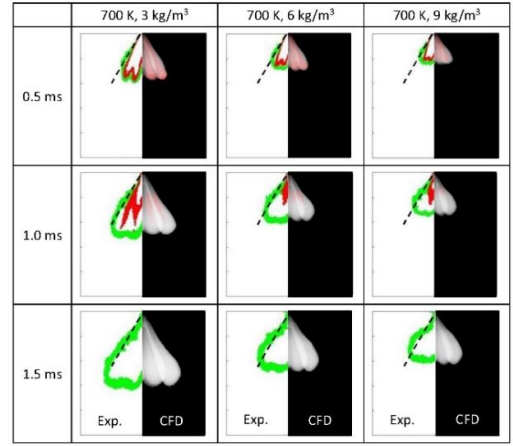

*Figure 6 Comparisons of experiment [10] and simulations for liquid and vapor envelopes at 700 K conditions.*

The liquid and vapor penetrations are plotted as line graphs in Figure 7. The solid lines are the experimental results, while the dashed lines are the CFD predictions and the colors indicate the different ambient densities. Similar to the observations in the 2D spray envelopes, the transient penetration of the vapor plume is well predicted by the model predictions with less than 5% error at 3.5 ms ASI. Quasi-steady state of the liquid length is reached after an initial transient period for both the 700 K and 900 K cases. However, such a quasi-steady period is not seen at 500 K, indicating that quasi-steady state of the liquid length can not be reached within such a short duration of injection at low temperature conditions. Even though the Liquid-Jet model used in the EP spray model is derived based on experiments of long-duration, quasi-steady state fuel injections, its application in the present CFD simulations is seen to capture the transient behavior of a short-duration injection very well, as shown in the predictions of liquid penetration for the 6 and 9 kg/m³ conditions.

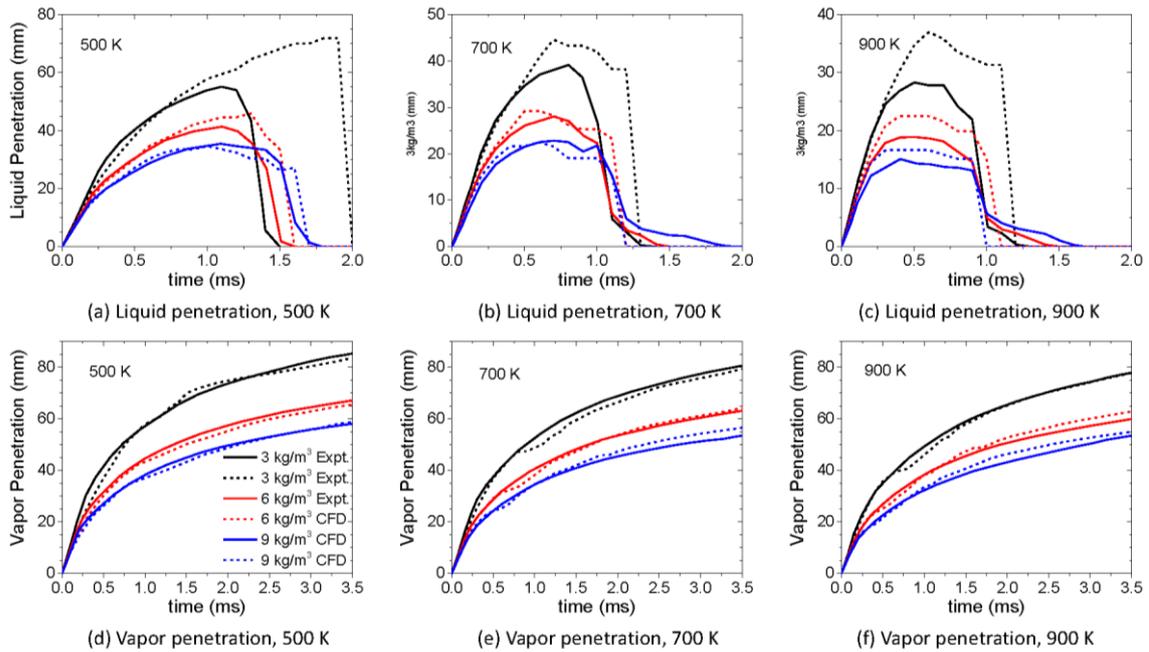

*Figure 7 Liquid and vapor penetrations. (a), (b) and (c) are liquid penetrations for 3, 6, 9 kg/m³ ambient densities, respectively; (d), (e) and (f) are their corresponding vapor penetrations. Black, red and blue lines correspond to 500 K, 700 K and 900 K ambient temperatures, respectively. Solid lines are experiments, and dashed liens are CFD predictions.*

Figure 8 shows the maximum liquid length as a function of ambient temperature for each ambient density condition. Excellent agreements can be found between the experimental results and the CFD predictions for most cases. The error in the prediction of the maximum liquid length grows with decreasing ambient density and temperature. In this case, the validity of the mixing-controlled assumption is believed not to be the reason of such error, as the phase equilibrium, mixing-controlled vaporization process would be expected to always give a faster vaporization and shorter liquid length than the non-equilibrium vaporization process. Possibly the constant value of $C_L$ in Equation 4 is questionable. $C_L$ has a theoretical value of 0.38 from the derivation of Siebers' scaling law, and was adjusted to 0.62 in order to match n-hexadecane and heptamethylnonane liquid length data for ambient temperatures from 700 K to 1300 K and ambient densities from 14.8 kg/m³ to 59.0 kg/m³ [9]. The change in $C_L$ is considered to be a compensation for errors introduced by assumptions made when deriving the liquid length scaling law, which should not be expected to be the same when the operating condition change significantly.

Moreover, the accuracy in the correlation for the spreading angle θ can also affect the liquid length prediction. Particularly, liquid cavitation within the nozzle passages is a possible agency in the determination of $C_\theta$ in Equation 5 [23], which is not considered in this study. The cavitation number K is usually used to quantify the cavitation transition, which is defined as $K = (P_{inj} - P_{amb})/(P_{amb} - P_v)$. $P_{inj}$ is the injection pressure, $P_{amb}$ is the ambient pressure, and $P_v$ is the fuel vapor pressure. For the studied cases in this section, the value of K ranges from 7.5 for the highest ambient pressure condition, to 70.4 for the lowest ambient pressure condition, which is estimated using $P_v$ of 0.78 bar for iso-octane at 363 K. A higher K value indicates higher tendency of cavitation, and consequently wider spreading angle. Thus, consideration of the K factor could mitigate the deviation in liquid length prediction seen at low pressure conditions. Nonetheless, the current form of the EP spray model provides satisfying results for most engine-relevant conditions, especially for the vapor plume, which is essential in engine combustion simulations.

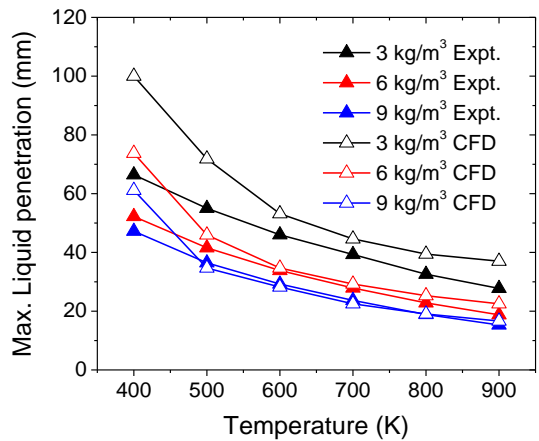

*Figure 8 Maximum liquid penetrations as function of ambient temperature. Black, red, blue lines correspond to 3, 6, 9 kg/m³ ambient densities, respectively. Solid squares are experimental results; hollow squares are CFD predictions.*

*Multi-component gasoline surrogate simulation*

Gasoline and diesel fuels are complex mixtures that consist of a number of paraffins, cycloalkanes and olefins, etc. While a single component surrogate is sometime used in experimental and numerical research of IC engine combustion for simplicity, the use of a multi-component surrogate that considers both light-end and heavy-end components is favored to represent the complicated vaporization behavior of gasoline and diesel fuels. Because the present phase equilibrium solver is capable of dealing with multi-component mixtures, the EP spray model can be extended to simulate multi-component fuel injection. To demonstrate this ability, a 10-component gasoline surrogate fuel [24] was used for the ECN Spray G condition. All the simulation setups were unchanged except for the fuel surrogate. The composition of the multi-component surrogate is listed in Table 1, along with properties for each species.

Figure 9 shows the 2D distribution of mixture fraction and mass fraction for iso-pentane, a light-end species, n-heptane, a species with moderate boiling temperature and m-cymene, a heavy-end species, at 0.7 ms ASI. The distribution patterns are found to be significantly different, depending on the fuel properties. Attributed to the lowest boiling temperature, iso-pentane quickly evaporates into the vapor phase and concentrates upstream and in the cold core of the plume. To the opposite, m-cymene shows a low vapor concentration in these cold regions and remains in the liquid phase due to its high boiling temperature, and only evaporates at the periphery of the plume where the temperature is relatively high due to ambient gas entrainment. N-heptane shows a distribution pattern somewhere between the two extreme cases of iso-pentane and m-cymene, which is similar to the overall mixture fraction distribution that considers all components.

*Table 1 10-component surrogate for 91 RON gasoline [24]*

| Component name | Chemical formula | Molecular weight [g/mol] | Boiling temperature at 1 bar [K] | Critical temperature [K] | Critical pressure [bar] | Mass fraction [-] |
|---|---|---|---|---|---|---|
| n-heptane | $n\text{-}C_7H_{16}$ | 100 | 371.6 | 540.2 | 27.4 | 0.05921 |
| n-decane | $n\text{-}C_{10}H_{22}$ | 142 | 447.2 | 617.7 | 21.1 | 0.06023 |
| 2,2,3,3 tetra-methylhexane | $C_{10}H_{22}$ | 142 | 413.5 | 623 | 25.1 | 0.02935 |
| iso-pentane | $i\text{-}C_5H_{12}$ | 72 | 301.2 | 460.4 | 33.8 | 0.31702 |
| iso-heptane | $i\text{-}C_7H_{16}$ | 100 | 363.2 | 530.4 | 27.4 | 0.13954 |
| iso-octane | $i\text{-}C_8H_{18}$ | 114 | 372.5 | 543.8 | 25.7 | 0.09062 |
| toluene | $C_7H_8$ | 92 | 383.8 | 591.8 | 41.1 | 0.09932 |
| m-xylene | $C_8H_{10}$ | 106 | 412.2 | 617.1 | 35.4 | 0.07041 |
| m-cymene | $C_{10}H_{14}$ | 134 | 448.2 | 657 | 29.3 | 0.10484 |
| 1-hexene | $C_6H_{12}$ | 84 | 336.2 | 504 | 32.1 | 0.02888 |

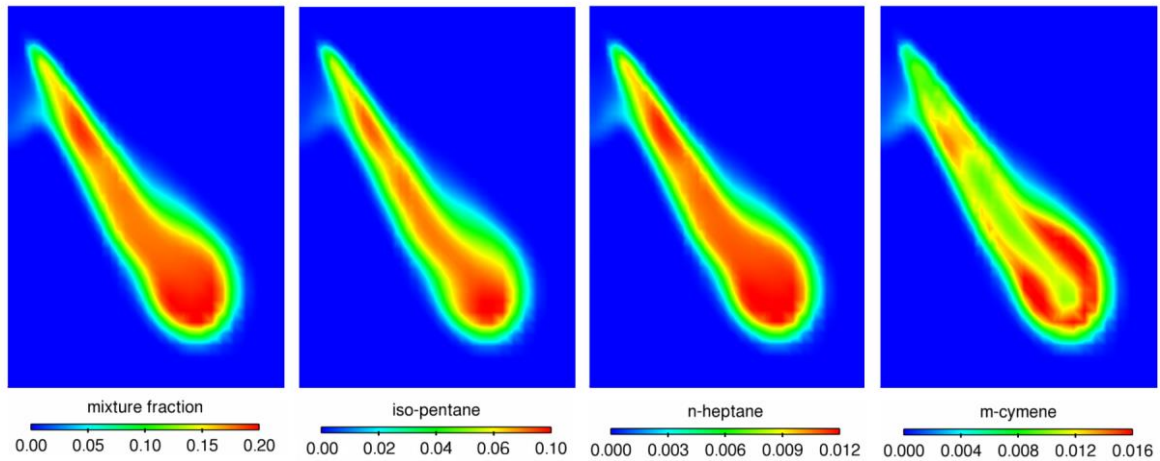

*Figure 9 Distribution of mixture fraction and mass fraction for different species.*

## Summary and Conclusions

In this study, the Equilibrium Phase (EP) spray model is applied to GDI spray simulations. The model is validated extensively under conditions from 3 kg/m$^3$ to 9 kg/m$^3$ and 400 K to 900 K, which covers the situations of interest to GDI engines from cold start, early injection, to late injection near TDC for charge stratification. The model shows good accuracy in vapor penetrations and plume envelope predictions for all the simulated conditions, and also in the liquid penetrations for most of the conditions from medium to high ambient density. The accuracy of liquid length prediction at low ambient densities could potentially be improved by using a comprehensive spray cone angle model that considers in-nozzle flow, cavitation, etc. Compared to diesel simulations, use of the EP model for GDI spray simulation requires the near-nozzle region to be appropriately spatially resolved due to the narrow included angle and close periphery between plumes. A 10-component gasoline fuel surrogate is also applied in the present simulations to demonstrate the capability of the EP model to simulate multi-component fuels. The component distribution is shown to be sensitive to the species properties, which could have significant impact on charge stratification, ignition and emission formation in engine combustion processes.

## Acknowledgements

This work was finished during the Ph.D project of the corresponding author, Yue, Z., at the University of Wisconsin-Madison. The authors would like to acknowledge the financial support provided by the China Scholarship Council (CSC). The authors are also thankful for support from ANSYS for use of EnSight software.